\documentclass{article}
\usepackage{spconf,amsmath,graphicx}
\usepackage[hidelinks]{hyperref}

\title{Framework for evaluation of sound event detection in web videos}

\name{Rohan Badlani$^{\star \ddag}$  \qquad Ankit Shah$^{\dagger \ddag }$ \thanks{$^{\ddag}$ First two authors contributed 
equally} \qquad Benjamin Elizalde \thanks{$^{\mathsection}$Acknowledges CONACYT for his doctoral fellowship, No.343964}  $^{\dagger \mathsection}$  \qquad Anurag Kumar$^{\dagger} \qquad $ Bhiksha Raj$^{\dagger}$  
}
\address{$^{\dagger}$ Language Technologies Institute, Carnegie Mellon University, Pittsburgh PA \\ $^{\star}$ Department of Computer Science, BITS Pilani, India \\ }

\begin{document}
\maketitle
\begin{abstract}
The largest source of sound events is web videos. Most videos lack sound event labels at segment level, however, a significant number of them do respond to {\em text  queries}, from a match found using metadata by search engines. In this paper we explore the extent to which a search query can be used as the true label for detection of sound events in videos. We present a framework for large-scale sound event recognition on web videos. The framework crawls videos using search queries corresponding to 78 sound event labels drawn from three datasets. The datasets are used to train three classifiers, and we obtain a prediction on 3.7 million web video segments. We evaluated performance using the search query as true label and compare it with human labeling. Both types of ground truth exhibited close performance, to within 10\%, and similar performance trend with increasing number of evaluated segments. Hence, our experiments show potential for using search query as a preliminary true label for sound event recognition in web videos.
\end{abstract}

\section{Introduction}
\label{sec:intro}
\vspace{-0.1in}
The Internet is being flooded with massive amount of multimedia data, mostly comprising of videos containing sound events which are often critical to understand the video content. Hence, it is necessary to automatically recognize sound events within the audio, {\em e.g.} \textit{police siren}, \textit{dishwasher} or \textit{birds singing}. Sound event recognition has been applied to multiple forms, such as in conjunction with other modalities to retrieve and index  consumer-generated videos based on content~\cite{yu2014informedia, jiang2010columbia, lan2012double,cheng2012sri}, video surveillance (e.g. detection of footsteps)~\cite{atrey2006audio}, human-robot interaction (e.g. detection of choke)~\cite{maxime2014sound,janvier2012sound}, wildlife monitoring (e.g. detection of animals)~\cite{ruiz2015multiple}, and context-aware systems (e.g. outdoors or home)~\cite{eronen2006audio}.  

In recent years, the main sound recognition challenges: DCASE 2013~\cite{DCASE2016workshop}, 2016 \cite{giannoulis2013detection} and 2017\cite{DCASE2017challenge} have fostered research providing standard datasets, task guidelines, metrics and benchmark performances. Although necessary, the literature tends to focus on DCASE-like datasets, which are audio-only recordings and smaller in scale. This leaves the primary source of sound events, the web, and its intrinsic problems less explored which makes it unclear how state of the art sound event recognition systems work on web videos. Although the YouTube based AudioSet~\cite{audioset2017} was recently released containing weak labels for sound events, our work explores mismatch conditions between existing audio-only research and datasets applied to YouTube videos.

Sound event recognition on large-scale web videos poses several challenges, mainly the lack of annotated audio recordings for sound events to train and evaluate systems. In order to exploit web recordings, unsupervised solutions have been explored, such as clustering~\cite{salamon2015unsupervised} and sound diarization~\cite{elizalde2012there} or semi-supervised approaches~\cite{han2016semi,shah2016approach} or visual domain web video analysis ~\cite{7780475,45493}, which learn from a combination of labeled and unlabeled data sources. Another technique~\cite{kumar2016audio} relies on weak labels for learning where only the presence or absence of sounds in the recording is known. For web videos the primary idea is that associated metadata can be used to assign weak labels which can further be used for training classification models \cite{hershey2017cnn,kumar2017deep}. However, the metadata, such as title, keywords and description, are noisy and often related to the visual information rather than the audio content, hence it remains to be seen how reliably it can be used as a true label or ground truth to train and evaluate sound recognition systems. This analysis forms the major contribution of the framework proposed in this paper. 

In this paper, we first do an exploration to identify the extent to which search query, which relates to the textual metadata, can be used as a true label for sound events at segment level for YouTube videos. This study, to the best of our knowledge is unavailable in the literature. For our study, we developed a framework for large-scale sound event recognition on web videos consisting of three modules, \emph{Crawl, Hear, Feedback}. In \emph{Crawl}, YouTube videos were crawled using search queries corresponding to 78 sound event labels and the keyword \textit{sound} ($<$sound event label$>$ sound) drawn from three datasets. In \emph{Hear}, the datasets are used to train three multi-class classifiers, which are used to obtain sound event label prediction on 3.7 million video segments. We evaluated performance using the search query as the true label and compare it on a subset against human labeling which was collected in the \emph{Feedback} module. Both types of ground truth exhibit similar performance trend. Hence, we show that search query provides a reasonable ground truth for large-scale sound event detection in web videos. 
\vspace{-0.1in}

\section{Framework}
\label{sec:frmwk}
\begin{figure}[h]
   \centering
     \includegraphics[width=0.5\textwidth]{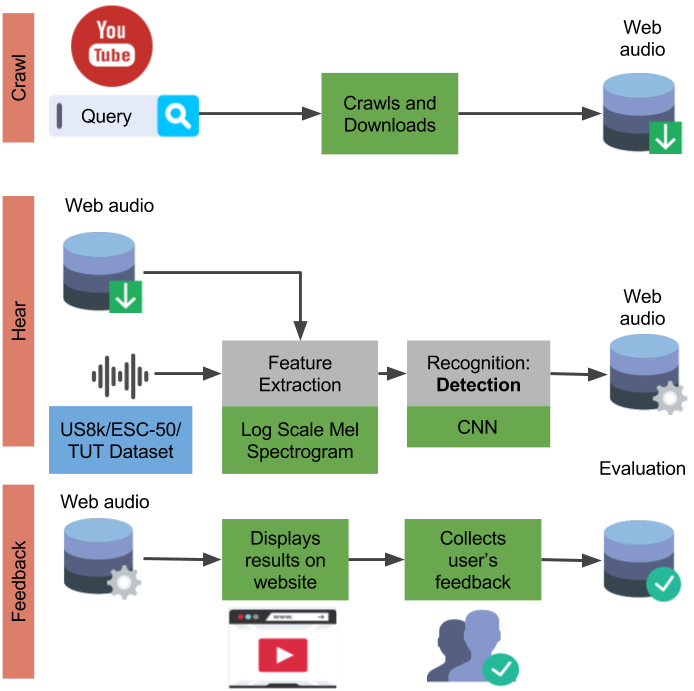}
       \vspace{-2mm}
     \caption{Our framework consists of three modules: Crawl, Hear and Feedback.}
     \vspace{-3mm}
     \label{fig:Diagram-Self-learning}
\end{figure}

\vspace{-0.1in}
The purpose of the framework is to use our sound event labels as search queries to crawl videos, which lack true labels at segment level; train classifiers using labeled audio to recognize sound events on the unlabeled crawled video segments; and evaluate the system performance using two types of ground truth, search query and human labeling collected through our website. The framework as described in following sections consists of three modules illustrated in Figure \ref{fig:Diagram-Self-learning}

\vspace{-0.1in}

\subsection{Crawl}
\label{crawl_fmwk}
\vspace{-0.1in}
The \textit{Crawl} module employs search queries to scrape audio from YouTube videos using the Pafy API\footnote{https://pypi.python.org/pypi/pafy}. The queries are kept to use them later as true labels.
\vspace{-0.1in}

\subsection{Hear}
\vspace{-0.1in}
\textit{Dataset Aggregator} organizes different annotated sound event datasets. The audio is then preprocessed and acoustic features are extracted in the \textit{Feature Extractor} module. We run a \textit{Sound Event Classifier}, the features are used to train classifiers, on unlabeled segments of the crawled videos. The performance is evaluated using the previously used search query and with the human inspection carried on the next module.   

\subsection{Feedback}
\vspace{-0.1in}
This module displays the classifier predictions along with corresponding audio segments on our website \textbf{\href{nels.cs.cmu.edu}{nels.cs.cmu.edu}}. Using our website, human feedback is collected [assumed as true label for an experiment] to evaluate classifier performance and compare the performance against search query as true label.
\vspace{-0.1in}

\section{Experiments and Evaluation}
\label{expt}
\vspace{-0.1in}
In this section, we explain how we use our framework to study the relation between the search query and the presence of sound events in video segments. To achieve our objective we trained sound event classifiers using labeled recordings sourced from three different audio-only datasets and run trained detectors on unlabeled crawled YouTube video segments. The performance was evaluated using two types of true labels - with the search query used to retrieve the videos and with the collected human inspection. 
\vspace{-0.11in}

\subsection{Crawl}\label{crawl_data_evaluation}
\vspace{-0.1in}
In contrast to audio-only recordings, collecting audio from videos poses several challenges. YouTube contains massive amount of videos and a proper formulation of the search query is necessary to filter videos with higher chances of containing the desired sound event. Typing a query composed by a noun such as \emph{air conditioner} will not necessarily fetch a video containing such sound event because the associated metadata often corresponds to the visual content; contrary to audio-only websites such as \emph{freesounds.org}. Therefore, we modified the query to be a combination of keywords: ``\textless sound event label\textgreater \hspace{0.1cm}sound'', for example,``air conditioner sound''. Although the results empirically improved, the sound event was not always found to be occurring and even if it was present, sometimes it was present within a short duration. We discarded videos longer than ten minutes and shorter than three seconds because they were either likely to contain unrelated sounds or were too short to be processed.

The defined search query was used to crawl videos corresponding to 78 sound event labels described in the following Section~\ref{Annotated_Datasets}. Around 260 hours of video was processed, equally distributed per audio event, which corresponds to over 3.7 million video segments (90\% overlap) of 2.3 seconds each. The segments were converted to 16-bit encoding, mono-channel, and 44.1 kHz sampling rate WAV files. Note that, the acoustic content from these videos is unstructured and the target sound is often overlapping by other audio, such as noise, speech or music. 
\vspace{-0.11in}

\subsection{Hear} \label{Experiment_Training}
\vspace{-0.1in}
In this subsection, we explain how we used three labeled datasets to train our three sound event classifiers and run them on the unlabeled crawled YouTube video segments. 

\vspace{-0.1in}
\subsubsection{Dataset Aggregator} \label{Annotated_Datasets}
\vspace{-0.1in}
The 78 sound events come from 3 publicly available annotated datasets - ESC50, US8k and TUT. We partitioned each dataset into 60\% training, 20\% validation and 20\% testing sets to avoid dealing with the costly process of cross-fold validation during testing of 3.7 million segments. 

\emph{ESC-50 or Environmental Sound Classification} ~\cite{piczak2015environmental} has 50 classes from five categories: animals, natural soundscapes and water sounds, human non-speech sounds, interior/domestic sounds and exterior sounds. ESC-50 consists of 2,000 audio segments with an average duration of 5 seconds. 

\emph{The US8K or UrbanSounds8K}~\cite{Salamon:UrbanSound:ACMMM:14} has 10 classes: \emph{air conditioner, car horn, children playing, dog bark, street music, gun shot, drilling, engine idling, siren, jackhammer.} UrbanSounds8k consists of 8,732 audio segments with an average duration of 3.5 seconds.

\emph{TUT 2016}~\cite{Mesaros2016_EUSIPCO} has 18 classes like \emph{car passing by, bird singing, door banging} from two major sound contexts namely home context and residential area. TUT dataset consists of 954 audio segments with an average duration of 5 seconds. 

\vspace{-0.1in}
\subsubsection{Feature Extractor} \label{Classifier_feature_extraction}
\vspace{-0.1in}
We extracted features for all audio recordings in the datasets based on the work in ~\cite{piczak2015environmental} because they provided near to state-of-the-art performance at the time of developing our experimental results. Our pipeline is agnostic of the classifier used. The audio recordings were re-sampled into 16-bit encoding, mono channel at 44.1 kHz sampling rate as a standard format for all experiments. We feed in two channels to our learning model. The first channel comprises of log-scaled mel-spectrograms with 60 mel-bands with a window size of 1024 (23 ms) and hop size is 512 and the second channel comprises of delta coefficients for mel-spectrograms. 

\vspace{-0.1in}
\subsubsection{Sound Event Classifiers} \label{Classifier_training}
\vspace{-0.1in}
We used multi-class classifiers using Convolutional Neural Networks (CNNs) for each of the datasets based on the work in~\cite{piczak2015environmental}. Thus, we trained 3 CNN models for classification of 50, 18 and 10 sound events from ESC-50, TUT and US8k respectively. We used different models for each dataset because using a single model for 78 audio events presented many challenges like dealing with unbalanced classes, inconsistency in feature normalization and doing so resulted in low performance (15\% lower accuracy).

The CNN architecture consisted of the following layer parameters and optimizations done using the validation set. The input to the CNN is $60 \times 101 \times 2$. We used 60 mel-filters, 101 number of frames (approximately 2.3 seconds of data) and 2 channels - mel-spectra and delta features for mel-spectrograms. The input window length of 101 frames is moved by 10 frames ($90\%$ overlap). Hence, we trained and predicted on audio segments of approximately $2.3$ secs. The first convolutional ReLU layer consisted of 80 filters of rectangular shape (57x6 size, 1x1 stride) allowing for slight frequency invariance. Max pooling was applied with a pool shape of 4x3 an stride of 1x3. A second convolutional ReLU layer consisted of 80 filters (1x3 size, 1x1 stride) with max pooling (1x3 pool size, 1x3 pool slide). Further processing was applied through two fully connected hidden layers of 5000 neurons with ReLU non-linearity. The final output layer is a softmax layer. Training was performed using Keras implementation of mini batch stochastic gradient descent even with shuffled sequential batches (batch size 1000) and a nestrov momentum of 0.9. We used L2 weight decay of 0.001 for each layer and dropout probability of 0.5 for all layers.

\vspace{-0.1in}

\subsection{Evaluation of Classifiers Performance}\label{web_data_evaluation}
\vspace{-0.1in}
YouTube videos at segment level lack of true labels for sound events. Hence, we evaluated the classification performance with two types of references or ground truth - the search query used to retrieve the videos and the human inspection collected with the website described in the \emph{Feedback} module.

\vspace{-0.1in}
\subsubsection{Evaluation assuming Search Query as Ground Truth} \label{query}
\vspace{-0.1in}
In this evaluation process, all the segments of a retrieved video using a given search query, such as \emph{dog barking sound} are labeled to contain \emph{dog barking}, even if this might not necessarily be true. Motivation is that search query is a reflection of the accumulated metadata tags such as title, description and keywords and hence, we wanted to see to what degree the query relates to the acoustic content of the segments.

\vspace{-0.1in}
\subsubsection{Evaluation using Human Feedback as Ground Truth}\label{human}
\vspace{-0.1in}
Human inspection is needed to provide the most reliable ground truth (true label). Hence, the 3.7 million predicted segments were sorted based on classifier confidence (probability) and were evaluated by a group of experts on tasks related to sound recognition. The top 40 segments for each of the 78 classes were distributed randomly among 5 human evaluators and at least 3 people evaluated each segment to reduce human bias and decide based on majority vote. The segments were displayed using a similar web interface as in the main page of \href{nels.cs.cmu.edu}{nels.cs.cmu.edu}, but the difference is that only the audio was displayed in lieu of video in order to avoid revealing other cues, such as images or title. The evaluators had to choose between two options, \textit{Correct} or \textit{Incorrect}, whether the evaluator claims that the system's predicted class was present within the segment or not.

\section{Results and Discussion}
\vspace{-0.1in}
\subsection{Results on Crawled YouTube Videos}
\begin{figure}[h]
  \centering
  \includegraphics[width=1\linewidth]{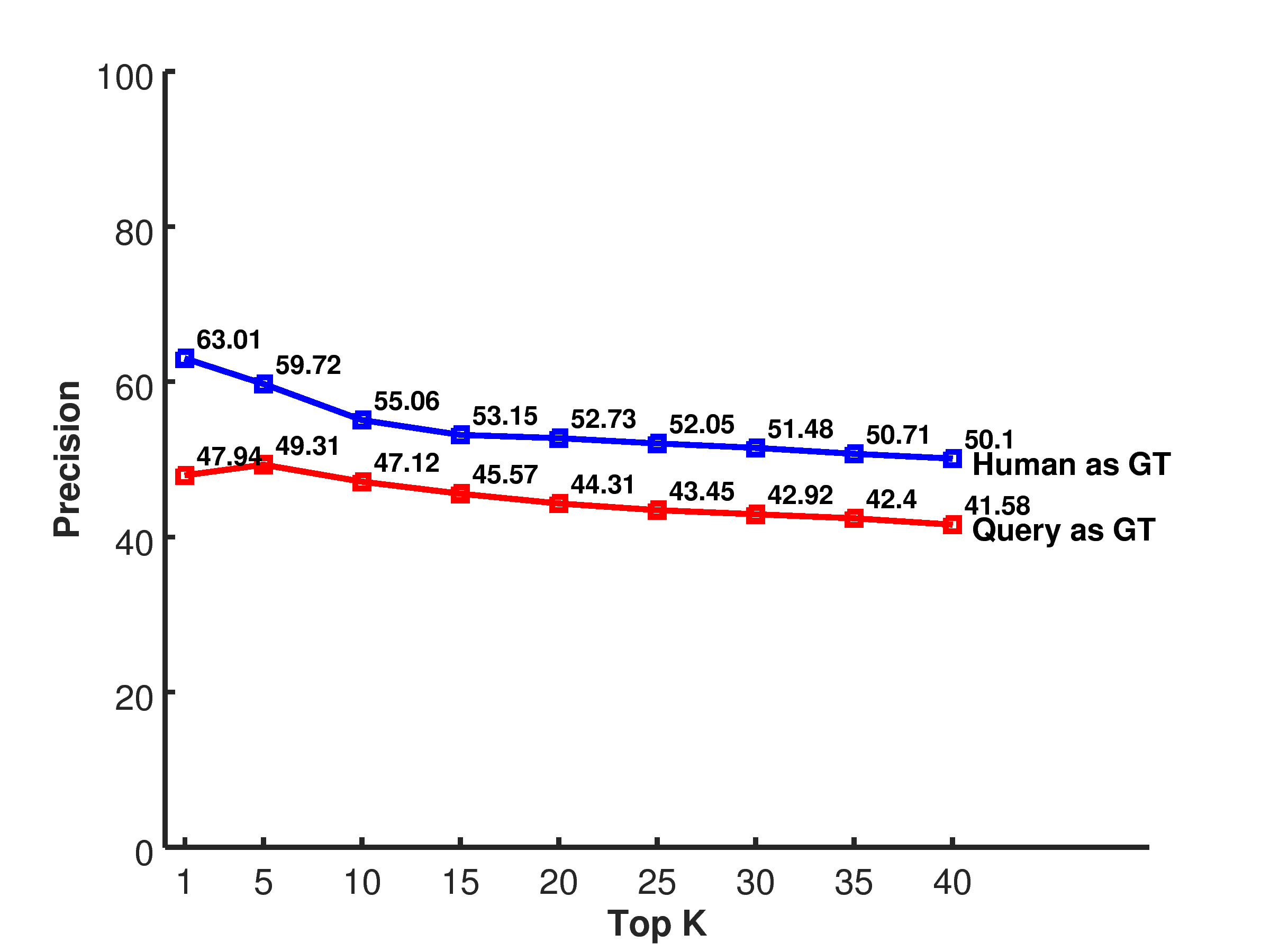}
  \vspace{-5mm}
  \caption{Performance for the combination (weighted average) of the three classifiers. (GT = ground truth)}
  \vspace{-6mm}
  \label{fig:Top_K_precision}
\end{figure}

\begin{figure}[h]
  \centering
  \includegraphics[width=0.95\linewidth]{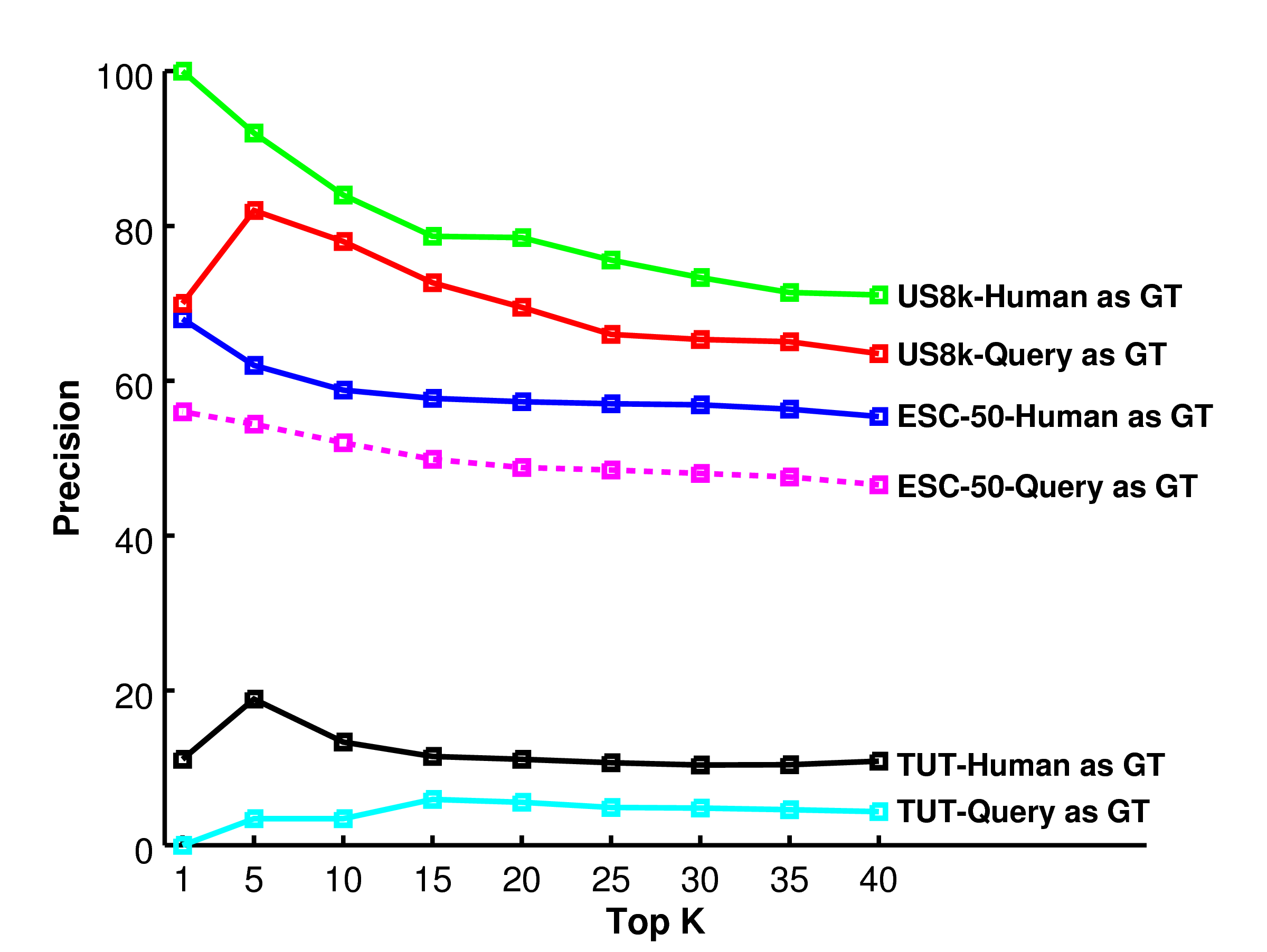}
    \vspace{-5mm}
  \caption{For the three classifiers, the search query-based performance follows a similar and close trend to the one based on human feedback. (GT = ground truth)}
  \vspace{-6mm}
  \label{fig:Dataset-topk}
\end{figure}

\vspace{-0.1in}
This main takeaway of our study is the exhibited correlation between the presence of sound events in video segments and their corresponding search query (including the keyword sound), illustrated in Figures \ref{fig:Dataset-topk} and \ref{fig:Top_K_precision}. Note that human inspection is the most reliable ground truth while the search query is an assumption of true class because it is based on metadata, which may be based on visual content. Thus, the precision with human feedback was expected to be higher than the one with the query, but it was uncertain how big that gap could be. The query-based performance was better than we expected considering the uncertainty of the audio content in web videos. Moreover, the performance follows a similar trend to the one from human feedback with a relatively close precision of less than an absolute 10\%, which shows potential for the search query (including the keyword sound) to be used as the class label in lieu of human annotations. 

The performance of the three classifiers on the video segments evaluated for both types of ground truth, search query and human feedback, are shown in both Figures \ref{fig:Top_K_precision} (combined weighted-average) and \ref{fig:Dataset-topk} (individual performance). The y-axis, has the performance in terms of Precision@K (a common retrieval metric), which is the precision of \textit{k} high-confidence (probability) ranked segments. The x-axis has the Top \textit{K} high-confidence segments yielded by our systems. In both figures, the results for K = 1-5 is unstable as the number of audio segments is small and could vary depending on the selected audio segments, however performance stabilizes as \textit{K} grows. We stopped at Top 40 results because a YouTube user, for example, tend to focus on the home page of results which translates to \textit{K} equals to 10-20. Further, we evaluated all the 3.7 million segments using search query as ground truth and obtained precision scores of 15.43\% for ESC-50, 33.58\% for US8k and 7.43\% for TUT datasets. Future work involves using crowd-sourcing to collect more human feedback to determine whether performance based on human inspection would remain within 10\% precision.
\vspace{-0.1in} 

\vspace{-0.1in}
\subsection{Results on Datasets}\label{sec:annotated_results}
\begin{figure}[h]
\centering
\includegraphics[width=0.5\textwidth]{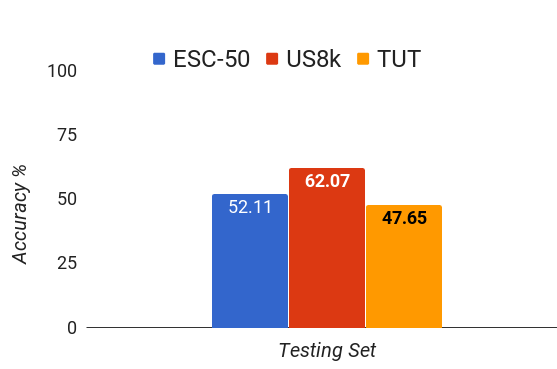}
\vspace{-6mm}
\caption{Classification accuracy for each of the three classifiers trained on the each of the three datasets showed reliable results on match conditions.}
\vspace{-5mm}
\label{fig:Detector_accuracy}
\end{figure}

\vspace{-0.1in}
The classification accuracy of our three systems on their corresponding testing sets is shown in order to establish their reliable performance in match conditions. Although the three datasets are well explored in the field, we split them in a different manner to avoid cross-fold experiments with 3.7 million segments. The classification accuracy for ESC-50 52.11\%, US8k 62.07\% and TUT 47.65\% was considerably better than their corresponding random performance: 2\%, 10\%, 5.5\% as shown in Figure \ref{fig:Detector_accuracy}.
\vspace{-0.1in}

\vspace{-0.1in}
\section{Conclusions} 
\vspace{-0.1in}
Web videos have no established true sound event labels at segment level. Thus, we studied the relation between search queries, based on sound event labels, and the presence of the corresponding sound event. We developed a framework to crawl videos using search queries, trained classifiers with audio-only datasets on video segments and evaluate performance with two types of ground truth - the search query and the collected human inspection. They showed a correlation between the search query (including the keyword sound) and the presence of sound events in video segments. Our results encourage further exploration of the search query as a preliminary label of the true class to evaluate sound event classification at a large-scale. 

\newpage
\ninept
\bibliographystyle{IEEEbib}
\bibliography{IEEE}

\end{document}